\documentclass[pre, nofootinbib, floatfix, notitlepage]{revtex4-1}
\usepackage[utf8]{inputenc}

\usepackage{amssymb,amsthm,amsmath, amsfonts}
\usepackage{color, graphicx, enumerate}
\usepackage[font={scriptsize}]{caption}
\usepackage{verbatim}

\begin{document}
\author{Katarzyna Macieszczak
}
\affiliation{University of Nottingham, School of Mathematical Sciences, School of Physics \& Astronomy, University Park, NG7 2RD Nottingham, UK}
\title{Upper bounds on the quantum Fisher Information\\ in the presence of general dephasing}
\begin{abstract}
We derive upper bounds on the quantum Fisher information in interferometry with $N$ subsystems, e.g. two-level atoms or Gaussian modes, in the presence of arbitrarily correlated Gaussian dephasing including independent and collective dephasing. The derived upper bound enables us to analyse the Fisher information asymptotic behaviour when $N\rightarrow\infty$. Dephasing introduces random phases to subsystems dynamics, which lowers the precision of estimating the phase difference $\phi$ in an interferometer. The method presented uses Bayesian estimation of the random phases and eliminates dephasing noise by calculating their weighted arithmetic mean, which correponds to the phase $\phi$ estimated in~interferometry.
\end{abstract}
\maketitle

\textbf{Introduction.}  In numerous areas of modern Physics, e.g. spectroscopy in atomic clocks~\cite{spectroscopy} or gravitational interferometers~\cite{GI}, it is necessary to estimate an unknown value of a parameter of quantum system dynamics. 

When using a quantum system, the precision of parameter estimation is bounded, from below, by the inverse of the Fisher information which depends on parameter encoding details, available resources, such as number $N$ of atoms/Gaussian modes in the system and initial system state preparation, and the measurement performed on the system state with an encoded parameter. When a parameter is encoded via the unitary dynamics of the system, the estimation error scaling can be improved from the classical shot-noise scaling $\propto N^{-1}$ to the Heisenberg scaling $\propto N^{-2}$ by an entangled initial state~\cite{GLM}. This quantum enhancement in precision, however, may be significantly limited in~the presence of decoherence, i.e. when the system interacts with an uncontrolled enviroment ~\cite{EMD},~\cite{G}. 

\begin{figure}[htb!]
\includegraphics[width=0.5\textwidth]{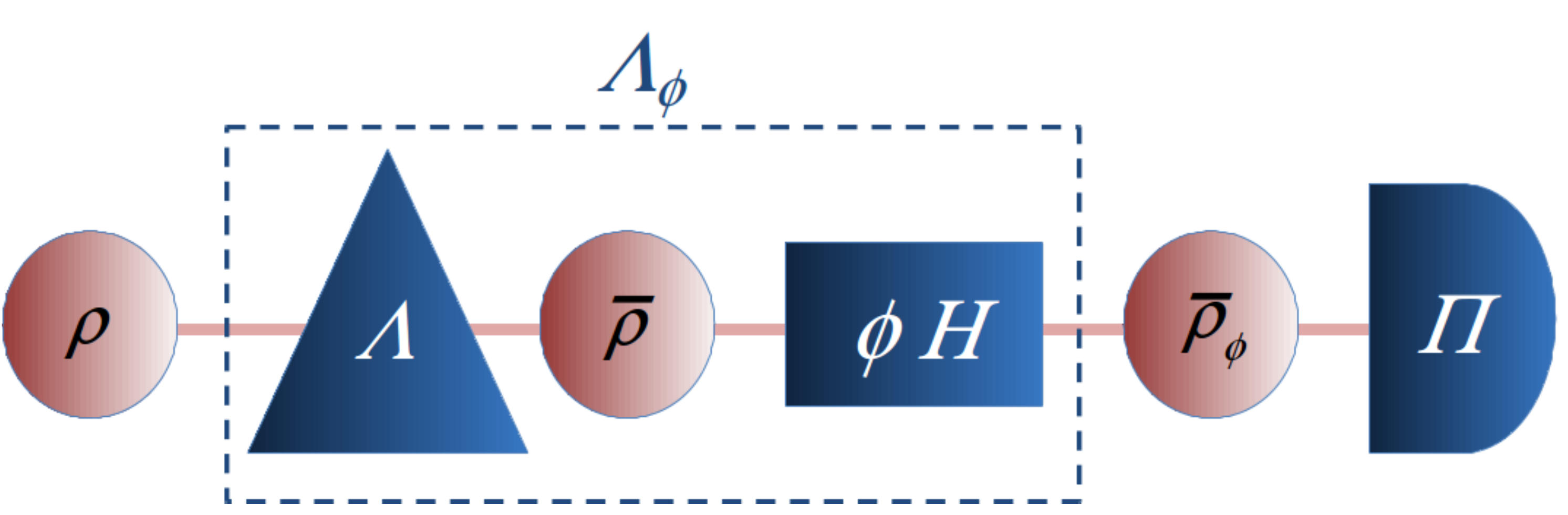}
\caption{The interferometry setup discussed in the paper: The parameter value $\phi$ is encoded in an initial state $\rho$ via a quantum channel $\Lambda_{\phi}$. A~dephasing channel $\Lambda$ which commutes with unitary encoding of $\phi$ is considered. The interferometer is described by a Hamiltonian $H$ which is the~generator of unitary encoding. A POVM measurement $\{\Pi_x\}_{x\in X}$ is performed on $\bar{\rho}_\phi$ to retrive the information about the value of $\phi$. Quality of the setup is quantified by the Fisher information.  \label{fig:scheme}} 
\end{figure}

The aim of interferometry is to estimate a phase $\phi$ encoded in an evolved system state, see Fig.~\ref{fig:scheme}. Here we discuss interferometry with $N$ subsystems, including two-level atoms or Gaussian modes, in the presence of dephasing.  Dephasing introduces additional random phases to subsystems dynamics, thus lowering the precision of the estimation of the phase $\phi$.  Only independent and collective dephasing have been successfully considered so far. Independent random phases have been discussed e.g. in~\cite{G} and the derived upper bound on Fisher information shows linear scaling with $N$. In~\cite{BE} the second case of identical random phases was considered and an upper bound, which converges to a constant when $N\rightarrow\infty$, was derived. We, however, provide a new unified approach to interferometry with $N$~subsystems, which delivers an upper bound on the Fisher information in the presence of arbitrarily correlated dephasing. This bound depends on both  correlationsof the random phase and initial system state preparation. It~is tight for weak decoherence, because the Heisenberg scaling of the estimation error is recovered as dephasing disappears. Furthermore, the bound enables us to analyse the  asymptotic behaviour of the Fisher information, thus~obtaining already familiar constant asymptotics for collective dephasing and the linear scaling in the independent case. \\ 

\textbf{Interferometry with dephasing.}  The interferometry setup is sketched out in Fig.~\ref{fig:scheme}. The system is first prepared in an initial state $\rho$ $(\rho\geq 0, \,\rho=\rho^{\dagger},\, \mathrm{Tr}\{\rho\}=1)$. Then it undergoes the dynamics described by a channel $\Lambda_\phi$ leading to an evolved state $\bar{\rho}_\phi=\Lambda_\phi(\rho)$ which has an enconded value of the $\phi$ phase. Finally, a POVM measurement $\{\Pi_x\}_{x\in X}$ ($\Pi_x\in \mathcal{B}(\mathcal{H})$, $\Pi_x\geq 0$, $\Pi_x=\Pi_x^{\dagger}$,  $\int_X\mathrm{d}x\,\Pi_x=1$) is performed on $\bar{\rho}_\phi$ in order to obtain information about $\phi$.
In the absence of dephasing, $\phi$ is encoded via unitary dynamics by a Hamiltonian $H$, i.e. the evolved state is~$\rho_\phi:=e^{-i\phi H }\rho\, e^{i \phi H}$. When interaction with an enviroment leads to dephasing, we have $\Lambda_\phi(\rho)=e^{-i\phi H }\Lambda(\rho)\,e^{i\phi H }$, where $\Lambda$ represents the dephasing channel, which commutes with unitary dynamics generated by $H$, see Eq.~(\ref{eq:state}). Thus,~dephasing can  also be viewed as an imperfect preparation of the initial state $\rho$ such that in fact the dephased state $\bar\rho:=\Lambda(\rho)$ is prepared and used in the unitary interferometry setup. The estimation precision is lowered by dephasing, as all possible initial states are effectivily the mere outputs of the dephasing channel $\Lambda$. 

Let $H=\sum_{j=1}^N H_j$, where $H_j$ is a self-adjoint operator on $j$-th subsystem. Dephasing introduces random phases into the dynamics of the subsystems:
\begin{equation}
\bar{\rho}=\int_{\mathbb{R}^N}\mathrm{d}\widetilde{\varphi}_1...\mathrm{d}\widetilde{\varphi}_N\, g(\widetilde{\varphi}_1,...,\widetilde{\varphi}_N)\,\, e^{-i\sum_{j=1}^N \widetilde{\varphi}_j H_j }\rho\, e^{i\sum_{j=1}^N\widetilde{\varphi}_j H_j },\label{eq:state}
\end{equation}
where $g$ is the distribution of the random phases $\widetilde{\varphi}_j$, $j=1,..,N$.
One works with the averaged state $\bar{\rho}$ as it is not possible to access enviroment degrees of freedom in order to choose values of the random phases. The $\bar{\rho}$ state  is~influenced by correlations of the random phases. Usually $g$ is assumed to be Gaussian when it is fully determined by random phase means and a covariance matrix $C$. Without loss of generality, we assume the means to equal 0. Independent identically distributed phases decribe \textbf{independent dephasing} when fully correlated $\widetilde{\varphi}_j=\widetilde{\varphi}_k$, $1\leq j,k\leq N$ correspond to \textbf{collective dephasing}. 

 For a state $\rho$ of $N=1$ two-level atom we have $H=\frac{1}{2}\sigma^z$, where $\sigma^z$ is the Pauli matrix along the $z$-axis. For dephasing given by a Gaussian random phase with a variance $2\beta^2$ and~a mean~0, we obtain the following dephased state $\bar\rho$  (in the eigenbasis of $\sigma^z$):
\begin{equation}
\bar{\rho}=\left( \begin{array}{cc}
\rho_{00} & \rho_{01}\,e^{- \beta^2}   \\
\rho_{10}\,e^{- \beta^2} & \rho_{11} \end{array} \right),\qquad\mathrm{where}\qquad \rho=\left( \begin{array}{cc}
\rho_{00} & \rho_{01}  \\
\rho_{10} & \rho_{11} \end{array} \right).
\end{equation}\\

\textbf{Fisher information.} The quality of the interferometry setup can by quantified using the Fisher information. 

Here, it is enough to discuss the case when $\phi$ is unitarily encoded in an initial state $\rho$, i.e. $\rho_\phi:=e^{-i\phi H }\rho\, e^{i \phi H}$, and~a~POVM measurement is performed on $\rho_\phi$. In the presence of dephasing we simply replace $\rho$ by $\bar\rho$.

We have to estimate $\phi\in\mathbb{R}$ only from a result $x$ of the POVM measurement $\{\Pi_x\}_{x\in X}$ performed on a state from the family $\{\rho_\phi\}_{\phi\in\mathbb{R}}$.  A result $x\in X$ is obtained with probability $p_{\phi}(x):=\mathrm{Tr}(\rho_\phi\Pi_x)$. In order estimate $\phi$ we use an~estimator - $\hat{\phi}:\,X\rightarrow\mathbb{R}$. Let us consider the case of $\phi=\phi_0+\delta\phi$, where $\phi_0$ is known and $\delta\phi\ll1$ is a small fluctuation that we want to estimate. We compare estimators by using the local error defined as $\Delta^2_{\phi_0}\hat{\phi}=\int_X \mathrm{d}x\, p_{\phi_0}(x)\, (\hat{\phi}(x)-\phi_0)^2$. For any locally unbiased estimator at $\phi=\phi_0$ ($\int_X\mathrm{d}x\,p_{\phi_0}(x)\hat{\phi}(x)=\phi_0$ and $\frac{\mathrm{d}}{\mathrm{d}\phi}|_{\phi=\phi_0}\int_X\mathrm{d}x\,p_{\phi}(x)\hat{\phi}(x)=1$) this error is~bounded from below in the Cramer-Rao inequality:
\begin{equation}
\Delta^2_{\phi_0}\hat{\phi}\geq F_{\phi_0,\rho, \Pi}^{-1}, \quad\mathrm{where}\quad F_{\phi,\rho, \Pi}=\int_{\{x\in X:\,p_{\phi}(x)\neq 0 \}}  \mathrm{d}x\, p_{\phi}(x)\,\left(\frac{\partial }{\partial \phi}\log(p_{\phi}(x))\right)^2  \label{eq:Fisher}
\end{equation}
where $F_{\phi, \rho, \Pi}$ is the Fisher information. This Fisher information quantifies the quality of the interferometry setup as~it bounds from below the phase estimation precision, thus simplifying the optimisation of the setup since we no loger need to refer to an estimator.

The Fisher information depends on the choice of measurement $\{\Pi_x\}_x\in X$. Whatever the measurement is~\cite{QFI}:
\begin{equation}
F_{\phi,\rho,\Pi}\leq F_{\rho_{\phi}}=\mathrm{Tr}(\rho_\phi L_{\rho_{\phi}}^2),\quad L_{\rho_{\phi}}\rho_{\phi}+\rho_{\phi}L_{\rho_{\phi}}=-i[H,\rho_\phi], \label{eq:qFisher}
\end{equation}
where $F_{\rho_{\phi}}$ is the quantum Fisher information and $L_{\rho_{\phi}}$ is the symmetric logarythmic derivative. The $L_{\rho_{\phi}}$  eigenbasis corresponds to the~optimal projective measurement for which $F_{\phi,\rho,\Pi}= F_{\rho_{\phi}}$.  As the quantum Fisher information is~the~same for all $\phi$, let us drop the index $\phi$. We have $F_{\rho}=\mathrm{Tr}(\rho L_\rho^2)$ and $L_{\rho}\rho+\rho L_{\rho}=-2i[H,\rho]$. Optimisation of~the~interferometry setup is reduced to choosing the initial state $\rho$.\\

It is not easy to find the maximum of the quantum Fisher information w.r.t. the initial state in the presence of~dephasing, even numerically. In order to discuss the asymptotic behaviour of~the~quantum Fisher information in~the~presence of~general dephasing, we need to derive a new upper bound. Our method uses knowledge about the~dephasing channel $\Lambda$, i.e. the random phase probability distribution. First, we estimate random phase values using the Bayesian approach. Then, we eliminate dephasing noise by calculating a weighted arithmetic mean of the~random phase estimators, which corresponds to $\phi$. This mean can be related to the optimal estimator of $\phi$ which~saturates the~Cramer-Rao bound in Eq.~(\ref{eq:Fisher}). Our approach provides a clear and simple insight into the quantum Fisher information behaviour in the presence of general dephasing. \\




\textbf{The Bayesian approach.} Let us consider one run of the intererometry experiment. First, unknown random phase values are chosen according to a Gaussian distribution in with all means equal $0$ and a covariance matrix $C$. Next, these phases are shifted by a common phase $\phi=\phi_0+\delta\phi$, where $\phi_0$ is known and $\delta\phi\ll1$ is an~uncontrolled fluctuation to~be~estimated. Let $\{\varphi_1,...,\varphi_N\}$ denote the shifted phase values and $g_{\phi}$ their probability distribution with all~ the~means now equal $\phi$, and the unchanged covariance matrix $C$: $g_{\phi}(\varphi_1,...,\varphi_N)=(2\pi\det C)^{-\frac{1}{2}} \exp(-\frac{1}{2}\sum_{j,k=1}^N(\varphi_j-\phi)(C^{-1})_{jk}(\varphi_k-\phi))$.  These values are then encoded in the initial state $\rho_{\varphi^{(N)}}= e^{-i\sum_{j=1}^N \varphi_j H_j }\rho\, e^{i\sum_{j=1}^N\varphi_jH_j }$ and we go on to perform the POVM measurement $\{\Pi_x\}_x\in X$. In order to estimate $\phi$ from a result $x\in X$,
we first estimate the shifted random phases $\varphi^{(N)}$ and then their common mean which equals exactly $\phi$. As we cannot choose the random phase value, in many experiments we obtain a result $x$ with an average probability $\bar{p}_{\phi}(x)=\int_{\mathbb{R}^N}\mathrm{d}\varphi^{(N)} \, g_{\phi}(\varphi^{(N)}) p_{\varphi^{(N)}}(x)=\mathrm{Tr}(\bar\rho_{\phi}\Pi_x)$, where $\bar{\rho}_{\phi}$ is the dephased state in Eq.~(\ref{eq:state}) and~$p_{\varphi^{(N)}}(x):=\mathrm{Tr}\{\rho_{\varphi^{(N)}}\Pi_x\}$. We would expect the Fisher information to appear, since $\delta\phi$ is small.

We know the $g_{\phi}$ distribution except for the mean $\phi$, which we need to estimate. Let us consider the following \emph{Gedankenexperiment}. We assume that we can observe phases $\varphi^{(N)}$ directly. In order to estimate $\phi$ we eliminate random dephasing noise by calculating a weighted arithmetic mean $\hat{\phi}(\varphi^{(N)}):= \sum_{j=1}^N\gamma_j\varphi_j$ with $\sum_{j=1}^N \gamma_j=1$, thus guaranteeing that $\hat{\phi}$ is an unbiased estimator of $\phi$. Using e.g. Lagrange multipliers one can show that $\gamma_j:=\frac{\sum_{k=1}^N( C^{-1})_{jk}}{\sum_{j,k=1}^N( C^{-1})_{jk}}$ leads to minimum local estimation error $\Delta^2_{\phi_0}\hat{\phi}:=\int_{\mathbb{R}^N}\mathrm{d}\varphi^{(N)}\,g_{\phi_0}(\varphi^{(N)})\,\left(\hat{\phi}(\varphi^{(N)})-\phi_0\right)^2=\left(\sum_{j,k=1}^N( C^{-1})_{jk}\right)^2=:\Delta_C^2$. Let~us~note that it is sufficient to measure just one phase $\varphi_C:=\sum_{j=1}^N\gamma_j\varphi_j$  in order to estimate $\phi$. For independent dephasing we have $\gamma_j=\frac{1}{N}$, $j=1,...,N$, and  $\Delta^2_C=\frac{2\beta^2}{N}$, where $2\beta^2:=C_{11}$ is the variance of every random phase.

We cannot, however, observe phases directly, but only via a measurement result $x\in X$. In order to estimate the~shifted phases we use knowledge about their Gaussian distribution $g_{\phi}$ and the Bayesian estimation. The Bayesian approach provides the estimators $\hat{\varphi}_j(x):=\frac{\int_{\mathbb{R}^N}\mathrm{d}\varphi^{(N)} \,g_{\phi}(\varphi^{(N)} )\,p_{\varphi^{(N)} }(x) \,\varphi_j}{\int_{\mathbb{R}^N}\mathrm{d}\varphi^{(N)} \,g_{\phi}(\varphi^{(N)} )\,p_{\varphi^{(N)} }(x) }$, $j=1,...,N$, which have minimum error w.r.t.~$g_{\phi}$ (see Appendix \ref{ap:Bayes}).  As we do not know the exact value of $\phi=\phi_0+\delta\phi$, we make an 'informed guess' assuming $\phi=\phi_0$ in order to obtain the random phase estimators. 

Inspired by our results for direct phase observation, we decided to take this a step further. We chose the estimator (abusing the notation) $\hat{\phi}(x):=\sum_{j=1}^N \gamma_j \hat{\varphi}_j(x)$ in order to find the $\phi$ value. This choice proved optimal up to a linear transformation which guarantees local unbiasedness at $\phi=\phi_0$ (see Appendix \ref{ap:optimal} for proof):
\begin{equation}
\hat{\phi}_{best}(x)=\phi_0+\frac{\hat{\phi}(x)-\phi_0}{\Delta_C^{-2}  \Delta^2_{\phi_0} \hat{\phi} }\quad\mathrm{and}\quad\Delta^2_{\phi_0}\hat{\phi}_{best}:=\int_X \mathrm{d}x\,\bar{p}_{\phi_0}(x) \,(\hat\phi_{best}(x)-\phi_0)^2=\left(\Delta_C^{-4} \Delta^2_{\phi_0} \hat{\phi}\right)^{-1}=F_{\phi_0,\bar{\rho},\Pi}^{-1}\label{eq:bestEST}
\end{equation}
as the Cramer-Rao inequality in Eq.~(\ref{eq:Fisher}) is saturated, which we prove as follows. We have $ \Delta_C^2\sum_{j=1}^N\frac{\partial}{\partial\varphi_j }g_\phi(\varphi_1,...,\varphi_N)=-g_\phi(\varphi_1,...,\varphi_N)\,(\varphi_C-\phi)$ bacuse of the definition of the $\varphi_C$ phase  and the fact that $g_\phi$ is Gaussian. We also have $\int_{\mathbb{R}^N}\mathrm{d}\varphi^{(N)} \,g_\phi({\varphi^{(N)}}) \sum_{j=1}^N \frac{\partial}{\partial\varphi_j} p_{\varphi_1,...,\varphi_N}(x) =\mathrm{Tr}\{-i[H,\bar{\rho}_\phi]\,\Pi_x\}$. Therefore:
\begin{eqnarray}
\hat{ \phi}(x)-\phi_0&:=&\frac{\int_{\mathbb{R}^N}\mathrm{d}\varphi^{(N)} \, g_{\phi_0}(\varphi^{(N)}) p_{\varphi^{(N)}}(x)\,(\varphi_C-\phi_0)\,}{\int_{\mathbb{R}^N}\mathrm{d}\varphi^{(N)} \, g_{\phi_0}(\varphi^{(N)}) p_{\varphi^{(N)}}(x)}=  \Delta_{C}^2\frac{\mathrm{Tr}\{-i[H,\bar{\rho}_{\phi_0}]\,\Pi_x\} }{\mathrm{Tr}\{\bar{\rho}_{\phi_0}\Pi_x\}}= \frac{\partial }{\partial \phi}|_{\phi=\phi_0}\log(\bar p_{\phi}(x))\quad \mathrm{and}\\ 
\Delta_{C}^{-4}\Delta^2_{\phi_0}\hat\phi&=&\Delta_{C}^{-4}\int_X \mathrm{d}x\,\bar{p}_{\phi_0}(x) \,(\hat\phi(x)-\phi_0)^2\,=\, F_{\phi_0,\bar{\rho},\Pi}. \label{eq:BayesFiGEN}
\end{eqnarray}
The relation in Eq.~(\ref{eq:BayesFiGEN}) was presented in a different context in~\cite{KM} in the case of one-dimensional Gaussian distribution, which can be related to collective dephasing.

In order to obtain the upper bound on the quantum Fisher information, let us look at the Bayesian estimation of the random phase $\varphi_C$, which has a Gaussian distribution $g_C$ with a mean $\phi$ and a variance $\Delta^2_C$. The above mentioned estimator $\hat{\phi}$ is also the best Bayesian estimator for phase $\varphi_C$ when $\phi=\phi_0$. Therefore, the average error of phase $\varphi_C$ estimation equals (see Appendix \ref{ap:Bayesmulti}): 
\begin{equation}
\Delta^2\hat{\phi}:=\int_{\mathbb{R}^N}\mathrm{d}\varphi^{(N)}\,g_{\phi}(\varphi^{(N)})\,\int_X\mathrm{d}x\, p_{\varphi^{(N)}}\left(\hat{\phi}(x)-\varphi_C\right)^2=\Delta_C^2-\Delta^2_{\phi_0}\hat\phi.\label{eq:BayesErrorMain}
\end{equation}
According to Eq.~(\ref{eq:BayesFiGEN}), the optimal measurements in the Bayesian estimation of $\varphi_C$ and in the Fisher information approach to $\phi$ estimation are exactly the same.

The average error $\Delta^2\hat{\phi}$ is bounded from below by the Bayesian Cramer-Rao inequality~\cite{BCR}. For a Gaussian distribution $g_C$ and the interferometry setup, we have (see Appendix \ref{ap:Bayesquantum}):
\begin{equation}
 \Delta^2\hat{\phi}\geq\left(\frac{1}{\Delta_{C}^2}+F_{\rho}\right)^{-1}, \label{eq:qBCR}
\end{equation}
where $F_\rho$ is the quantum Fisher information for the initial state $\rho$. Combining  Eqs.~(\ref{eq:BayesFiGEN}),~(\ref{eq:BayesErrorMain}) we obtain:
$F_{\phi_0,\bar{\rho},\Pi}\leq \left(\Delta_{C}^{2}+\frac{1}{F_\rho}\right)^{-1}  $.
We thus arrive at the \textbf{main result} of this paper, maximising $F_{\phi,\bar{\rho},\Pi}$ w.r.t. to the measurement:
\begin{equation}
F_{\bar{\rho}}\leq \left(\Delta_{C}^{2}+\frac{1}{F_\rho}\right)^{-1}, \label{eq:boundGEN}
\end{equation}
which, in turn, leads to the following bound on phase $\phi$ estimation precision for any locally unbiased estimator $\hat{\phi}$:
\begin{equation}
\Delta^2_{\phi_0}\hat{\phi}\geq \left(\Delta_{C}^{2}+\frac{1}{F_\rho}\right). \label{eq:boundGENerror}
\end{equation}
This bound, which takes into account both dephasing strength via $\Delta^2_C$ and the available resources via $F_{\rho}$, can be interpreted as follows. If a perfect random phase $\varphi_C$ observation were possible, the local error would be $\Delta_C^{2}$. As this is not possible, the error is greater by ${F_\rho^Q}$, taking into account the noise of observing phases only via the results of the interferometry experiment. 
We recover precision $F_\rho^{-1}$, which characterises the unitary dynamics, when the random phases variances converge to 0, since that implies $\Delta^2_C\rightarrow 0$. This guarantees that the bound will be tight in~the~presence of~weak dephasing. \\

\begin{figure}[htb!]
\begin{tabular}{cc}\includegraphics[width=0.5\textwidth]{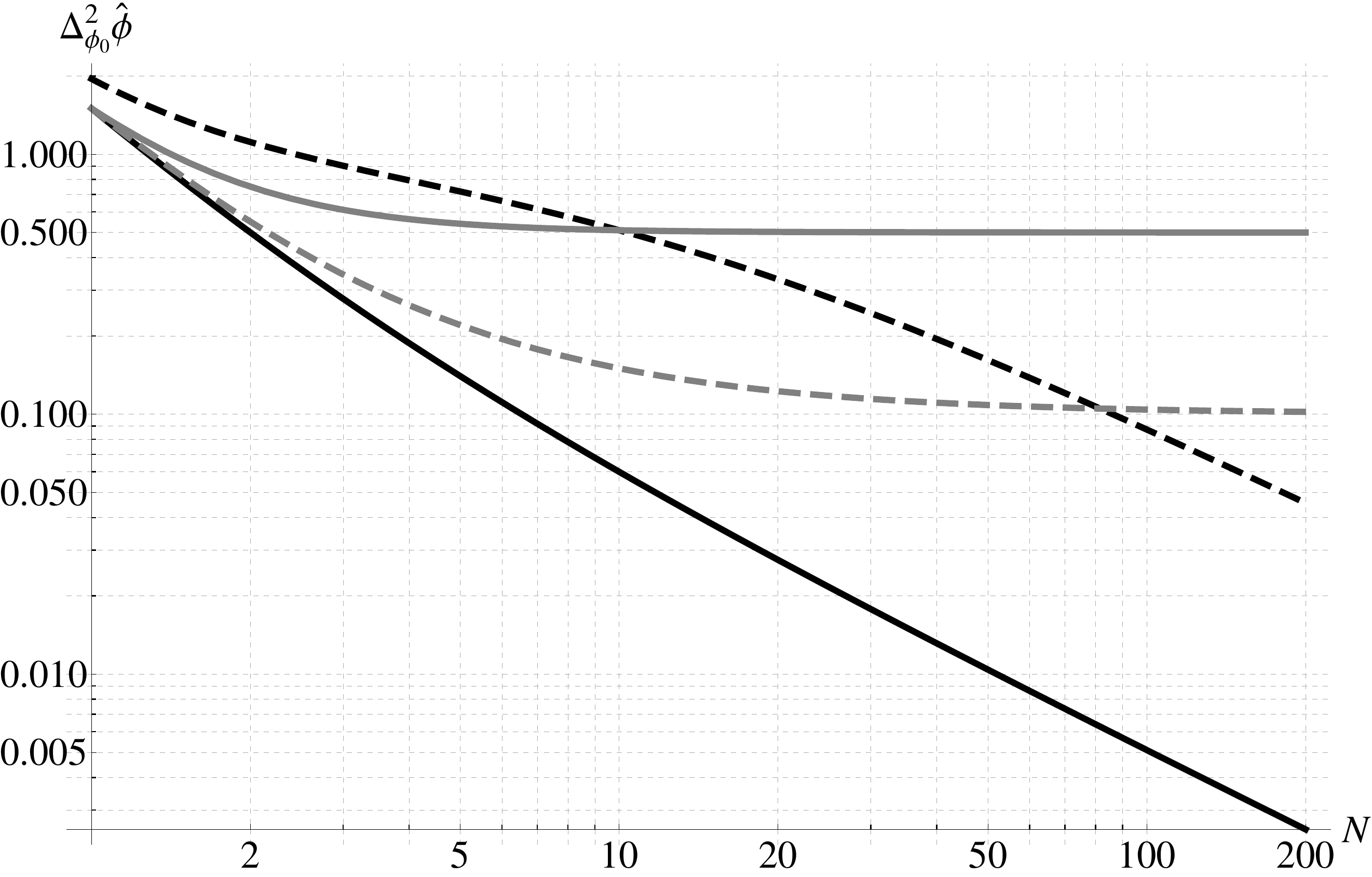}&
\includegraphics[width=0.5\textwidth]{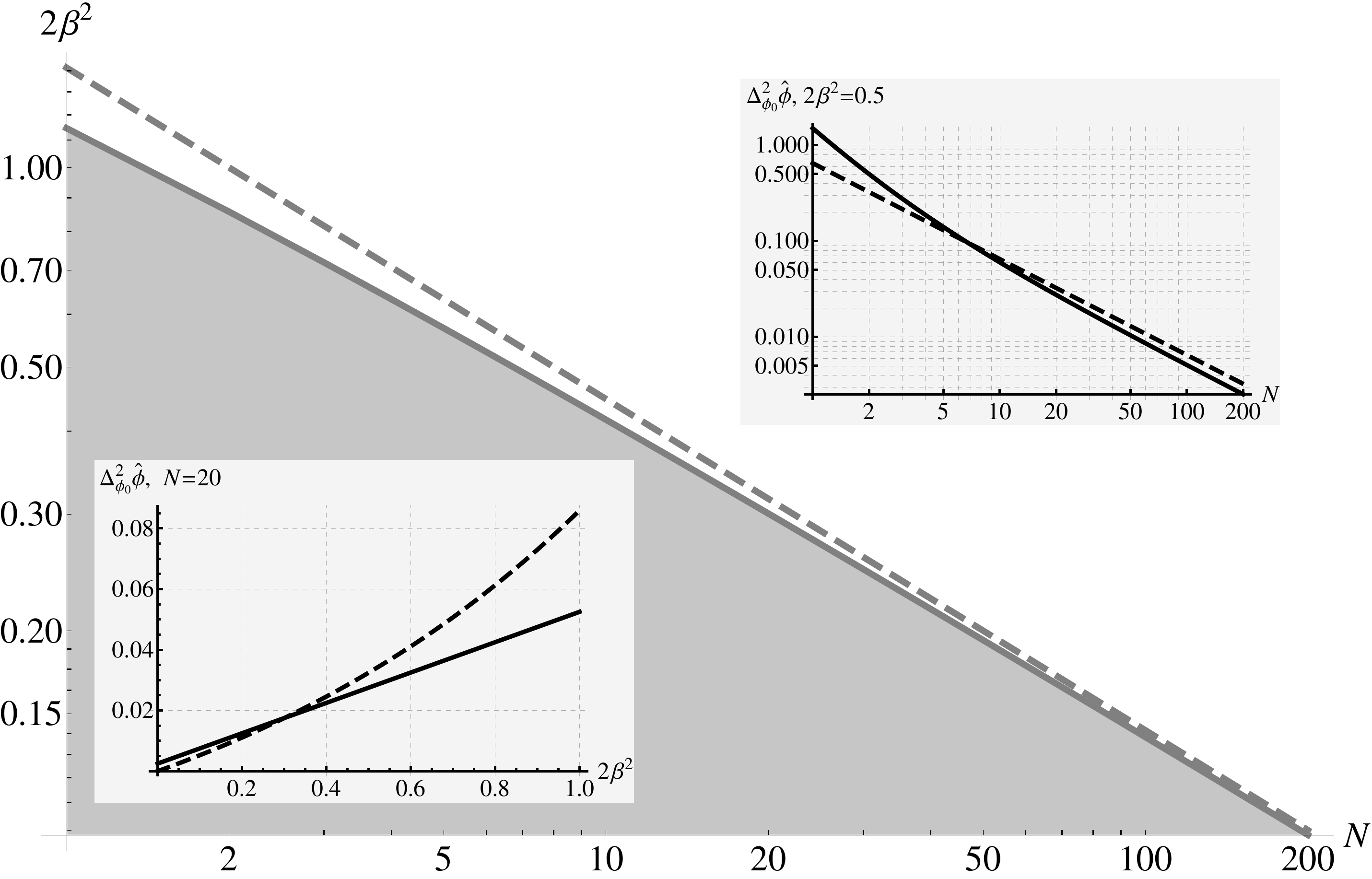}
\end{tabular} 
\caption{\textbf{RHS.} Bounds on the estimation error for $N$ two-level atoms obtained using Eq.~(\ref{eq:boundGENerror}). The difference in scaling with $N$, between independent $\propto N^{-1}$ (black solid line) and collective dephasing $\propto 1$ (gray solid line) is clearly visible. Weak (exponentially decaying) correlations in the $C_1$ example preserve the $\propto N^{-1}$ scaling (black dashed line; $\alpha=0.9$), whereas strong (non-decaying) correlations in the $C_2$ example limit the precision scaling to a constant error (gray dashed line; $\alpha=0.2$).   Dephasing strength $2\beta^2=0.5$ was chosen.\\ \textbf{ LHS.} Comparison of the bounds for $N$ two-level atoms in the presence of independent dephasing: the derived bound in Eq.~(\ref{eq:boundIND}) and the bound in~\cite{G}. The shaded area correponds to the values of $2\beta^2$ and $N$ for which the bound Eq.~(\ref{eq:boundIND}) is tighter (greater). When $N\rightarrow \infty$ this area is approximated by $2\beta^2=(2 N)^{-1/2}$ (gray dashed line). The insets depict the bound in Eq.~(\ref{eq:boundIND}) (solid line) and the bound in~\cite{G} (dashed line) w.r.t.  $2\beta^2$ (lower inset) or $N$ (upper inset). \label{fig:bounds}} 
\end{figure}

\textbf{Examples.} Let us consider the two following examples of correlated dephasing with the covariance matrices:
\begin{equation}
C_1=2\beta^2\left(\begin{array}{ccccc} 
1&\alpha&\multicolumn{2}{c}{\cdots}&\alpha\\
\alpha&1&\alpha&\cdots&\alpha\\
\vdots&\ddots&\ddots&\ddots&\vdots\\
\alpha&\cdots&\alpha&1&\alpha\\
\alpha&\multicolumn{2}{c}{\cdots}&\alpha&1\\
\end{array}\right)\quad\mathrm{and}\quad C_2=2\beta^2\left(\begin{array}{ccccc} 
1&\alpha&\alpha^2&{\cdots}&\alpha^{N-1}\\
\alpha&1&\alpha&\cdots&\alpha^{N-2}\\
\vdots&\ddots&\ddots&\ddots&\vdots\\
\alpha^{N-2}&\cdots&\alpha&1&\alpha\\
\alpha^{N-1}&\cdots&\alpha^2&\alpha&1\\
\end{array}\right).
\end{equation}
We obtain:
\begin{equation}
\Delta_{C_1}^{2}={2\beta^2}\left(\alpha+\frac{1-\alpha}{N} \right)\quad\mathrm{and}\quad\Delta_{C_2}^{2}=2\beta^2\,N^{-1} \frac{1+\alpha}{1-\alpha+\frac{\alpha}{N}}.\label{eq:Cexamples}
\end{equation}
We see that, for any value $0<\alpha<1$, for constant correlations (discrete topology) the bound in Eq.~(\ref{eq:boundGENerror}) converges to a constant $2\beta^2\,\alpha$, whereas for exponentially decaying correlations in one dimension, we obtain a better  asymptotic scaling $\thicksim 2\beta^2\,N^{-1} \frac{1+\alpha}{1-\alpha}\propto N^{-1}$, see the LHS in Fig. \ref{fig:bounds}. 

The case $\alpha=1$ corresponds to collective dephasing, both for $C_1$ and $C_2$:
\begin{equation}
\Delta^2_{\phi_0}\hat{\phi}\geq  \left(2\beta^2 +\frac{1}{F_\rho}\right) \label{eq:boundFC}
\end{equation}
which, for a single-mode Gaussian state $\rho$ of photons in a two-arm interferometer with an average number $\bar{N}$, due~to~$F_\rho\leq 8\bar{N}(\bar{N}+1)$ the form $
\Delta^2_{\phi_0}\hat{\phi}\geq   \left(2\beta^2 +\frac{1}{8\bar{N}(\bar{N}+1)}\right)$.
In~\cite{BE} this was proved using a variational approach to~the~quantum Fisher information. The bound in Eq.~(\ref{eq:boundFC}) has a \textbf{constant asymptotic behaviour}.
As the case $\alpha=0$ corresponds to independent dephasing, we obtain:
\begin{equation}
\Delta^2_{\phi_0}\hat{\phi}\geq N^{-1} \left(2\beta^2 +\frac{N}{F_\rho}\right).  \label{eq:boundIND}
\end{equation}
Given that for a state $\rho$ of $N$ two-level atoms we have $H=\frac{1}{2}\sum_{j=1}^N\sigma^z_j$, where $\sigma^z$ is the Pauli matrix along the $z$-axis, and thus  $F_\rho\leq N^2$, we arrive at
$\Delta^2_{\phi_0}\hat{\phi}\geq  N^{-1} \left(2\beta^2 +\frac{1}{N}\right)$.
In~\cite{G} a different upper bound $\Delta^2_{\phi_0}\hat{\phi}\geq  N^{-1} (e^{2\beta^2}-1)  $ was proved. The bounds are compared on the RHS in Fig. \ref{fig:bounds}. The bound in~\cite{G} works better for strong dephasing $\beta^2\geq (2N)^{-1,2}$, but does not provide Heisenberg scaling when $2\beta^2\rightarrow0$ . Both bounds show the scaling $\propto N^{-1}$ when~$N\rightarrow\infty$.

The bound in Eq.~(\ref{eq:boundGENerror}) provides an insight into interferometry in the presence of dephasing, the asymptotic precision of which is determined by the noise correlations. In the case of collective dephasing, if we were able to estimate the phases of atoms perfectly, it would be only one  phase $\varphi$ being a Gaussian variable with the variance $2\beta^2$. This is the exact bound  in Eq.~(\ref{eq:boundFC}) when~$N\rightarrow\infty$. If the phases are strongly correlated, as in the $C_2$ example, we effectively have a finite number of 'independent' noise realisations and cannot completely eliminate the dephasing noise, even~if~$N\rightarrow\infty$. Thus we observe that the bound in Eq.~(\ref{eq:boundGENerror}) converges to the constant $2\beta^2\alpha$.  If the phases are weakly correlated, as in the $C_1$ example, we can eliminate the noise, but the best possible scaling will be reduced from the Heisenberg scaling $\thicksim N^{-2}$ to the shot-noise scaling $\thicksim N^{-1}$.\\

\textbf{Summary and comments.} In this paper we present a new upper bound on the quantum Fisher information in~the~presence of arbitrarily correlated Gaussian dephasing that we have derived. This bound, as shown in Eq.~(\ref{eq:boundGEN}), takes into account both dephasing correlations and initial system state preparation. Moreover, it enables one to~analyse the asymptotic scaling of phase estimation precision when the number of subsystems $N\rightarrow\infty$. We also show that~weak (exponentially decaying) correlations of dephasing noise preserve the scaling $\propto N^{-1}$ which is characteristic in independent dephasing. Arbitrarily small, but strong (non-decaying) correlations limit the precision scaling when~$N\rightarrow\infty$ to a constant error. 

The bound derived can be fruitfully modified to frequency estimation~\cite{KM2}. \\

\\

\textbf{Acknowledgements.} The author is grateful to M\u{a}d\u{a}lin Gu\c{t}\u{a} and Sammy Ragy for inspiring discussions. We~would also like to thank Michael Hush for helpful comments on the draft of this work. This research was supported by~the~School of Mathematical Sciences and the School of Physics \& Astronomy at the University of Nottingham.

\begin{appendix}

\section{Single-parameter Bayesian estimation \label{ap:Bayes}} 

\textbf{Setup.} Let us assume that we know the probability distribution $g$ of a random variable $\varphi\in\mathbb{R}$.  Let $\phi$ denote the~mean and $\Delta^2$ the variance of the $g$ distribution. A value of $\varphi$  cannot be observed directly, but only via experiment results. We would like to estimate teh value of $\varphi$ from an experiment result $x\in X$, the probablity of which $p_\varphi(x)$ depends on $\varphi$. We look for an estimator $\hat{\varphi}:\,X\rightarrow \mathbb{R}$  with the smallest average error w.r.t. the $g$ distribution . 

For the average error  $\Delta^2\hat{\varphi}:=\int_{\mathbb{R}}\mathrm{d}\varphi\,g(\varphi)\int_X\mathrm{d}x\,p_\varphi(x)\,(\hat\varphi(x)-\varphi)^2 $ the optimal estimator is known to be: 
\begin{equation}
\hat{\varphi}(x):=\frac{\int_{\mathbb{R}}\mathrm{d}\varphi\,g(\varphi)\,p_\varphi(x)\,\varphi}{\int_{\mathbb{R}}\mathrm{d}\varphi\,g(\varphi)\,p_\varphi(x)}. \label{eq:BayesEST}
\end{equation}
In such a choice of estimator we have:
\begin{eqnarray}
\mathbb{E}\hat{\varphi}&:=&\int_{\mathbb{R}}\mathrm{d}\varphi\,g(\varphi)\int_X\mathrm{d}x\,p_\varphi(x)\,\hat\varphi(x) =\phi , \quad\mathrm{thus} \nonumber\\
\Delta^2\hat{\varphi}&=&\int_{\mathbb{R}}\mathrm{d}\varphi\,g(\varphi)\,(\varphi-\phi)^2 - \int_{\mathbb{R}}\mathrm{d}\varphi\,g(\varphi)\int_X\mathrm{d}x\,p_\varphi(x)\, (\hat\varphi(x)-\phi)^2  =  \Delta^2-\int_X\mathrm{d}x\,\bar{p}_\phi(x)\,(\hat\varphi(x)-\phi)^2,  \label{eq:BayesError}
\end{eqnarray}
where $\bar{p}_\phi(x):=\int_{\mathbb{R}}\mathrm{d}\varphi\,g(\varphi)\,p_\varphi(x)$ is average probability of obtaining the result $x\in X$. \\

\textbf{Bayesian Cramer-Rao bound}.  The Bayesian Cramer-Rao inequality bounds  from below  the average error $\Delta^2\hat{\varphi}$ of any estimator $\hat{\varphi}$~\cite{BCR}. For a Gaussian prior distribution $g$ with a variance $\Delta^2$ it is as follows:
\begin{equation}
\Delta^2\hat{\varphi}\geq\left(\frac{1}{\Delta^2}+\int_{\mathbb{R}}\mathrm{d}\varphi\,g(\varphi)F_{\varphi}\right)^{-1} \label{eq:BCR}.
\end{equation}
where $F_{\varphi}$ is the Fisher information for the $p_\varphi(\cdot)$ probability  defined as $F_{\varphi}:=\int_{\{x\in X:\,p_{\varphi}(x)\neq 0 \}}  \mathrm{d}x\, p_{\varphi}(x)\,\left(\frac{\partial }{\partial \varphi}\log(p_{\varphi}(x))\right)^2$, for a quantum setup see also Eq.~(\ref{eq:Fisher}).

\section{Reduction of multiparameter Bayesian estimation to single-parameter Bayesian estimation \label{ap:Bayesmulti}}

We are interested in estimating a random phase $\varphi_C=\sum_{j=1}^N\gamma_j\varphi_j$, where the random phases $\{\varphi_1,...,\varphi_N\}$ have a Gaussian distribution $g_\phi$ with a covariance matrix $C$ and the same means equal $\phi$, i.e. $g_\phi(\varphi_1,..,\varphi_N)=(2\pi\det C)^{-\frac{1}{2}}\exp(-\frac{1}{2}\sum_{j,k=1}^N(\varphi_j-\phi)(C^{-1})_{jk}(\varphi_k-\phi))$,  and $\gamma_j=\frac{\sum_{k=1}^N (C^{-1})_{jk}}{\sum_{j,k=1}^N (C^{-1})_{jk}}$, $j=1,...,N$. The distribution $g_C$ of~$\varphi_C$ is Gaussian with the variance equal $\Delta^2_C=\left(\sum_{j,k=1}^N (C^{-1})_{jk}\right)^{-1}$ and the mean equal $\phi$.

We cannot observe a $\varphi_C$ value directly, but only via an experiment result $x\in X$. The probability of obtaining a~result $x\in X$ $p_{\varphi_1,...,\varphi_N} (x)$ depends on all values $\{\varphi_1,...,\varphi_N\}$. The probability of obtaining $x\in X$ when $\varphi_C=\varphi$ is~$\int_{M_\varphi}g_\phi(\varphi_1,....,\varphi_N)\, p_{\varphi_1,...,\varphi_N} (x)=:p'_{\varphi}(x)$, where $M_\varphi:=\{\{\varphi_1,...,\varphi_N\}\in\mathbb{R}^N:\,\sum_{j=1}^N\gamma_j\varphi_j=\varphi\}$. 

The best estimator of $\varphi_C$ according to Eq.~(\ref{eq:BayesEST}) is: 
\begin{equation}
\hat{\varphi}_C(x):=\frac{\int_{\mathbb{R}}\mathrm{d}\varphi\,g_C(\varphi)\,p'_\varphi(x)\,\varphi}{\int_{\mathbb{R}}\mathrm{d}\varphi\,g_C(\varphi)\,p'_\varphi(x)}=\sum_{j=1}^N\gamma_j\hat{\varphi}_j(x), \label{eq:BayesESTC}
\end{equation}
where $\hat{\varphi}_j(x):=\frac{\int_{\mathbb{R}^N}\mathrm{d}\varphi_1...\mathrm{d}\varphi_N\,g_\phi(\varphi_1,....,\varphi_N)\,p_{\varphi_1,....,\varphi_N}(x)\,\varphi_j}{\int_{\mathbb{R}^N}\mathrm{d}\varphi_1...\mathrm{d}\varphi_N\,g_\phi(\varphi_1,....,\varphi_N)\,p_{\varphi_1,....,\varphi_N}(x)}$ is the best Bayesian estimator of a random phase $\varphi_j$ w.r.t. $g_\phi$. From Eqs.~(\ref{eq:BayesError}) and~(\ref{eq:BCR}) we arrive at:
\begin{eqnarray}
\mathbb{E}\hat{\varphi}_C=\phi,\quad \Delta^2\hat{\varphi}_C =\Delta_C^2-\int_X\mathrm{d}x\,\bar{p}_\phi(x)\,(\hat\varphi_C(x)-\phi)^2
\quad&\mathrm{ and}&\quad
\Delta^2\hat{\varphi}_C\geq\left(\frac{1}{\Delta_C^2}+\int_{\mathbb{R}}\mathrm{d}\varphi\,g_C(\varphi)F'_{\varphi}\right)^{-1}, \label{eq:BayesC}
\end{eqnarray}
where $\bar{p}_\phi(x):=\int_{\mathbb{R}}\mathrm{d}\varphi\,g_C(\varphi)\,p'_\varphi(x)=\int_{\mathbb{R}^N}\mathrm{d}\varphi_1...\mathrm{d}\varphi_N\,g_\phi(\varphi_1,....,\varphi_N)\,p_{\varphi_1,....,\varphi_N}(x)$ and $F'_{\varphi}$ is the Fisher information for~the~$p'_\varphi(\cdot)$ probability , i.e. $F'_{\varphi}:=\int_{\{x\in X:\,p'_{\varphi}(x)\neq 0 \}}  \mathrm{d}x\, p'_{\varphi}(x)\,\left(\frac{\partial }{\partial \varphi}\log(p'_{\varphi}(x))\right)^2$.

\section{Bayesian estimation in a quantum setup \label{ap:Bayesquantum}}

We perform a POVM measurement $\{\Pi_x\}_{x\in X}$ on a state $\rho_{\varphi_1,...,\varphi_N}:=e^{-i\sum_{j=1}\varphi_j\, H_j}\rho\,e^{i\sum_{j=1}\varphi_j\, H_j}$. The probability of obtaining a result $x\in X$ equals $p_{\varphi_1,...,\varphi_N} (x)=\mathrm{Tr}(\rho_{\varphi_1,...,\varphi_N}\Pi_x)$. When $\{\varphi_1,...,\varphi_N\}$ are Gaussian random variables with the same means equal $\phi$, we have $\bar{p}_\phi(x)=\mathrm{Tr}(\bar{\rho}_{\phi}\Pi_x)$, where $\bar\rho_\phi$ corresponds to the dephased state $\rho_\phi$, see Eq.~(\ref{eq:state}).\\

To use the Bayesian Cramer-Rao bound in Eq.~(\ref{eq:BCR}) we need the  Fisher information $F'_\varphi$ for the probability distribution $p'_\varphi(x):=\mathrm{Tr}(\rho'_\varphi \Pi_x)$, where $\rho'_\varphi$ is a state obtained by integrating $\rho_{{\varphi_1,...,\varphi_N} }$ over a set $\left\{{\varphi_1,...,\varphi_N} \in\mathbb{R}^N:\, \sum_{j=1}^N \gamma_j\varphi_j =\varphi \right\}$ with a conditional probility $g_\phi({\varphi_1,...,\varphi_N} |\varphi)=g_\phi({\varphi_1,...,\varphi_N} )/g_C(\varphi)$, where $g_C$ is the $\varphi_C$ probability distribution. 

Below we prove that $\frac{\mathrm{d}}{\mathrm{d}\varphi } \rho'_{\varphi}= -i\left[H,\rho'_{\varphi}\right]$, where $H=\sum_{j=1}^NH_j$.
Therefore, for $\rho':= e^{i\varphi \,H}\rho'_{\varphi}\,e^{-i\varphi \,H}$ we have  $F'_{\varphi}\leq F_{\rho'}$. As the quantum Fisher information is convex w.r.t. density matrices, we also have $ F_{\rho'}\leq F_{\rho}$. Thus, we~arrive at a quantum version of the Bayesian Cramer-Rao inequality above in Eq.~(\ref{eq:BayesC}):  
\begin{equation}
\Delta^2\hat{\varphi}_C\geq\left(\frac{1}{\Delta_{C}^2}+F_{\rho}^Q\right)^{-1}.
\end{equation}

We now prove that $\frac{\mathrm{d}}{\mathrm{d}\varphi } \rho'_{\varphi}= -i\left[H,\rho'_{\varphi}\right]$.  As  $\sum_{j=1}^N \gamma_j=1$, we obtain:
\begin{eqnarray}
\rho'_{\varphi}&=& g_C(\varphi)^{-1}\left(\sum_{j=1}^N\gamma_j\right)\int_{M_\varphi} g_\phi\left(\varphi_1,...,\varphi_N\right)\,\rho_{\varphi_1,...,\varphi_N}\nonumber\\
&=&g_C(\varphi)^{-1}\sum_{j=1}^N\gamma_j\int_{\mathbb{R}^{N-1}} \mathrm{d}\varphi_1...\mathrm{d}\varphi_{j-1} \mathrm{d}\varphi_{j+1}...\mathrm{d}\varphi_N\,g_\phi\left(\varphi_1,...,\varphi_{j-1},\gamma_j^{-1}\varphi-\gamma_j^{-1}\sum_{k=1,k\neq j}^N\gamma_k\varphi_k,\varphi_{j+1},...,\varphi_N\right)\, \nonumber\\
&&\qquad\qquad\qquad\times\,\rho_{\varphi_1,...,\varphi_{j-1},\gamma_j^{-1}\varphi-\gamma_j^{-1}\sum_{k=1,k\neq j}^N\gamma_k\varphi_k,\varphi_{j+1},...,\varphi_N},\nonumber\\
g_C(\varphi)&=&\sum_{j=1}^N\gamma_j\int_{\mathbb{R}^{N-1}} \mathrm{d}\varphi_1...\mathrm{d}\varphi_{j-1} \mathrm{d}\varphi_{j+1}...\mathrm{d}\varphi_N\,g_\phi\left(\varphi_1,...,\varphi_{j-1},\gamma_j^{-1}\varphi-\gamma_j^{-1}\sum_{k=1,k\neq j}^N \gamma_k\varphi_k,\varphi_{j+1},...,\varphi_N\right).
\end{eqnarray}
Therefore:
\begin{eqnarray}
 \frac{\mathrm{d}}{\mathrm{d}\varphi }g_C(\varphi)&=&\sum_{j=1}^N\gamma_j\int_{\mathbb{R}^{N-1}} \mathrm{d}\varphi_1...\mathrm{d}\varphi_{j-1} \mathrm{d}\varphi_{j+1}...\mathrm{d}\varphi_N\,g_\phi\left(\varphi_1,...,\varphi_{j-1},\gamma_j^{-1}\varphi-\gamma_j^{-1}\sum_{k=1,k\neq j}^N\gamma_k\varphi_k,\varphi_{j+1},...,\varphi_N\right)\nonumber\\ 
&&\qquad\qquad\times\,-\gamma_j^{-1}\left(  \sum_{k=1, k\neq j}^N (C^{-1})_{jk}(\varphi_k-\phi)\,+\, (C^{-1})_{jj} \left(\gamma_j^{-1}\varphi-\gamma_j^{-1}\sum_{k=1,k\neq j}^N\gamma_k\varphi_k-\phi\right)  \right)\nonumber\\
&=&-\sum_{j=1}^N \int_{M_{\varphi}} g_\phi(\varphi_1,...,\varphi_N)  \sum_{k=1}^N (C^{-1})_{jk} (\varphi_k-\phi)\,=\, - \int_{M_{\varphi}} g_\phi(\varphi_1,...,\varphi_N)  \sum_{k=1}^N (\varphi_k-\phi) \sum_{j=1}^N (C^{-1})_{jk} 
\nonumber\\
&=& - \int_{M_{\varphi}} g_\phi(\varphi_1,...,\varphi_N)  \sum_{k=1}^N (\varphi_k-\phi) \,\gamma_k \,\Delta^{-2}_C \,=\,-\int_{M_{\varphi}} g_\phi(\varphi_1,...,\varphi_N) \, (\varphi-\phi)\, \Delta_{C}^{-2}
\nonumber\\
& =&-\Delta^{-2}_C\,(\varphi-\phi) \,g_C(\varphi)
\end{eqnarray}
and
\begin{eqnarray}
\frac{\mathrm{d}}{\mathrm{d}\varphi } \rho'_{\varphi}&=&
g_C(\varphi)^{-1}\sum_{j=1}^N\gamma_j\int_{\mathbb{R}^{N-1}} \mathrm{d}\varphi_1...\mathrm{d}\varphi_{j-1} \mathrm{d}\varphi_{j+1}...\mathrm{d}\varphi_N\, g_\phi\left(\varphi_1,...,\varphi_{j-1},\gamma_j^{-1}\varphi-\gamma_j^{-1}\sum_{k=1,k\neq j}^N\gamma_k\varphi_k,\varphi_{j+1},...,\varphi_N\right)\nonumber\\ 
&&\qquad\qquad\times\,
\left[-i\gamma_j^{-1}H_j, \rho_{\varphi_1,...,\varphi_{j-1},\gamma_j^{-1}\varphi-\gamma_j^{-1}\sum_{k=1,k\neq j}^N\gamma_k\varphi_k,\varphi_{j+1},...,\varphi_N}\right]\nonumber\\
&+&  
g_C(\varphi)^{-1}\sum_{j=1}^N\gamma_j\int_{\mathbb{R}^{N-1}} \mathrm{d}\varphi_1...\mathrm{d}\varphi_{j-1} \mathrm{d}\varphi_{j+1}...\mathrm{d}\varphi_N\, g_\phi\left(\varphi_1,...,\varphi_{j-1},\gamma_j^{-1}\varphi-\gamma_j^{-1}\sum_{k=1,k\neq j}^N\gamma_k\varphi_k,\varphi_{j+1},...,\varphi_N\right)\nonumber\\ 
&&\qquad\qquad\times\,
\,-\gamma_j^{-1}\left(  \sum_{k=1, k\neq j}^N (C^{-1})_{jk}(\varphi_k-\phi)\,+\, (C^{-1})_{jj} \left(\gamma_j^{-1}\varphi-\gamma_j^{-1}\sum_{k=1,k\neq j}^N\gamma_k\varphi_k-\phi\right)  \right) \nonumber\\ 
&&\qquad\qquad\times\,\rho_{\varphi_1,...,\varphi_{j-1},\gamma_j^{-1}\varphi-\gamma_j^{-1}\sum_{k=1,k\neq j}^N\gamma_k\varphi_k,\varphi_{j+1},...,\varphi_N}\nonumber\\
&-&\rho'_{\varphi} g_C(\varphi)^{-1} \frac{\mathrm{d}}{\mathrm{d}\varphi }g_C(\varphi)
\nonumber\\
&=& -i\left[H,\rho'_{\varphi}\right]- \rho'_{\varphi} \Delta_{C}^{-2} (\varphi-\phi) + \rho'_{\varphi} \Delta_{C}^{-2} (\varphi-\phi)\,=\,  -i\left[H,\rho'_{\varphi}\right]\quad\blacksquare.
\end{eqnarray}

\section{Optimal locally unbiased estimator \label{ap:optimal}}
Let us prove that the choice of $\hat{\phi}_{best}(x):=\phi_0+\frac{\hat{\phi}(x)-\phi_0}{\Delta_C^{-2}  \Delta^2_{\phi_0} \hat{\phi}}$ in Eq.~(\ref{eq:bestEST}), where $\hat{\phi}(x):=\sum_{j=1}^N \gamma_j \hat{\varphi}_j(x)$ and $\hat{\varphi}_j$  is the best Bayesian estimator of the random phase $\varphi_j$ w.r.t. the $g_{\phi_0}$ distribution, $j=1,...,N$, is locally unbiased. Given that $\hat{\phi}_{best}$ saturates the Cramer-Rao inequality in Eq. (\ref{eq:Fisher}), the following will prove its optimality.

Let $\mathbb{E}_{\phi}$ denote average w.r.t. $\bar{p}_{\phi}(x)$. We have $\mathbb{E}_{\phi_0}\hat{\phi}=\phi_0$ from Eq.~(\ref{eq:BayesC}) and therefore also $\mathbb{E}_{\phi_0}\hat{\phi}_{best}=\phi_0$. We also have:
\begin{eqnarray}
\frac{\mathrm{d}}{\mathrm{d}\phi}|_{\phi=\phi_0}\mathbb{E}_{\phi}\hat{\phi}&=&\int_X\mathrm{d}x \,\frac{\mathrm{d}}{\mathrm{d}\phi}|_{\phi=\phi_0}\bar{p}_{\phi}(x)\,\hat{\phi}(x)=\int_X\mathrm{d}x \,\mathrm{Tr}\left(\frac{\mathrm{d}}{\mathrm{d}\phi}|_{\phi=\phi_0}\bar{\rho}_{\phi}\Pi_x\right)\,\hat{\phi}(x),\\
\frac{\mathrm{d}}{\mathrm{d}\phi}|_{\phi=\phi_0}\bar{\rho}_{\phi}&=&\int_{\mathbb{R}^N}\mathrm{d}\varphi_1... \mathrm{d}\varphi_N \,\frac{\partial}{\partial\phi}|_{\phi=\phi_0} g_{\phi}(\varphi_1,...,\varphi_N)\,\rho_{\varphi_1,...,\varphi_N}\nonumber \\
&=&\Delta^{-2}_C \int_{\mathbb{R}^N}\mathrm{d}\varphi_1... \mathrm{d}\varphi_N \,\left(\sum_{j=1}^N\gamma_j\varphi_j -\phi \right)g_{\phi_0}(\varphi_1,...,\varphi_N)\rho_{\varphi_1,...,\varphi_N},
\end{eqnarray}
where  $\Delta_C^2:=\left(\sum_{j,k=1}^N (C^{-1})_{jk}\right)^{-1}$ and $\gamma_j:=\Delta_C^2 \sum_{k=1}^N (C^{-1})_{jk} $, $j=1,..,N$.  Thus, since $\hat{\phi}(x):=\sum_{j=1}^N \gamma_j \hat{\varphi}_j(x)$  is also the best Bayesian estimator of $\varphi_C:=\sum_{j=1}^N \gamma_j \varphi_j$ w.r.t. $g_{\phi_0}$ we obtain:
\begin{eqnarray}
\frac{\mathrm{d}}{\mathrm{d}\phi}|_{\phi=\phi_0}\bar{p}_{\phi}(x)&=&\Delta^{-2}_C \,\bar{p}_{\phi_0}(x)\,(\hat{\phi}(x)-\phi_0)  \quad\mathrm{as} \quad\int_X\mathrm{d}x \,\frac{\mathrm{d}}{\mathrm{d}\phi}|_{\phi=\phi_0}\bar{p}_{\phi}(x)=0,\\
\frac{\mathrm{d}}{\mathrm{d}\phi}|_{\phi=\phi_0}\mathbb{E}_{\phi}\hat{\phi}&=&\int_X\mathrm{d}x \,\frac{\mathrm{d}}{\mathrm{d}\phi}|_{\phi=\phi_0}\bar{p}_{\phi}(x)\,(\hat{\phi}(x)-\phi_0)=\Delta_C^{-2}\int_X\mathrm{d}x \,\bar{p}_{\phi_0}(x)\,(\hat{\phi}(x)-\phi_0)^2=\Delta_C^{-2}\Delta^2_{\phi_0}\hat\phi,
\end{eqnarray}
which implies $\frac{\mathrm{d}}{\mathrm{d}\phi}|_{\phi=\phi_0}\mathbb{E}_{\phi}\hat{\phi}_{best}=1$, i.e. $\hat{\phi}_{best}$ is locally unbiased.

\end{appendix}

\end{document}